\documentstyle[12pt]{article}
\begin{document}

\baselineskip 7mm
\newcommand{\TeV}{\,{\rm TeV}}
\newcommand{\GeV}{\,{\rm GeV}}
\newcommand{\MeV}{\,{\rm MeV}}
\newcommand{\keV}{\,{\rm keV}}
\newcommand{\eV}{\,{\rm eV}}
\newcommand{\Tr}{{\rm Tr}\!}
\renewcommand{\arraystretch}{1.2}
\newcommand{\be}{\begin{equation}}
\newcommand{\ee}{\end{equation}}
\newcommand{\bea}{\begin{eqnarray}}
\newcommand{\eea}{\end{eqnarray}}
\newcommand{\ba}{\begin{array}}
\newcommand{\ea}{\end{array}} 
\newcommand{\bmat}{\left(\ba} 
\newcommand{\emat}{\ea\right)}
\newcommand{\refs}[1]{(\ref{#1})}
\newcommand{\ler}{\stackrel{\scriptstyle <}{\scriptstyle\sim}}
\newcommand{\ger}{\stackrel{\scriptstyle >}{\scriptstyle\sim}}
\newcommand{\lag}{\langle}
\newcommand{\rag}{\rangle}
\newcommand{\ns}{\normalsize}
\newcommand{\cm}{{\mathcal M}}
\newcommand{\gr}{m_{3/2}}
\newcommand{\p}{\partial}
\def\321{$SU(3)\times SU(2)\times U(1)$}
\def\21{$SU(2)\times U(1)$}   
\def\dbd{$0\nu\beta\beta~$}   
\def\ord{{\mathcal O}}
\def\tl{{\tilde{l}}}
\def\tL{{\tilde{L}}}
\def\bd{{\overline{d}}}   
\def\tL{{\tilde{L}}}
\def\a{\alpha}
\def\b{\beta}
\def\g{\gamma}
\def\c{\chi}
\def\d{\delta}
\def\D{\Delta}
\def\db{{\overline{\delta}}}
\def\Db{{\overline{\Delta}}}
\def\e{\epsilon}
\def\l{\lambda}
\def\n{\nu}
\def\m{\mu}
\def\nt{{\tilde{\nu}}}
\def\p{\phi}
\def\P{\Phi}
\def\solm{\Delta_{\odot}}
\def\sola{\theta_{\odot}} 
\def\mee{m_{ee}}
\def\atm{\Delta_{\makebox{\tiny{\bf atm}}}}
\def\k{\kappa}
\def\x{\xi}   
\def\r{\rho}
\def\mnu{{\mathcal M}_\nu}
\def\ml{{\mathcal M}_l}
\def\ue3{|U_{e3}|}
\def \soa{\frac{\Delta_{\odot}}{\Delta_{\makebox{\tiny{\bf atm}}}}}
\title{
\hfill hep-ph/0411154\\[.5cm]
Summary of Model Predictions for $U_{\lowercase{e}3}$ \footnote{Updated
version of the invited talk presented at the 5$^{th}$ Workshop on {\it
Neutrino Oscillations and their Origin (NOON2004)}, February
11-15,Tokyo, Japan}} \author{ Anjan S. Joshipura\\
{\ns\it ~ Theoretical Physics Group, Physical Research Laboratory,}\\
{\ns\it Navarangpura, Ahmedabad, 380 009, India.}\\}
\maketitle
\date{}
\begin{abstract}
We present a short discussion on the expected magnitude
of $|U_{e3}|$ in the context of various scenarios proposed to describe
neutrino masses and mixing. Generic expectation is relatively large
($>0.05$) values for $\ue3$ which occur in many well-motivated 
theoretical scenarios and models.
\end{abstract}
\section{Introduction}
The neutrino oscillations seen in the atmospheric and the solar  
neutrino experiments have been used to obtain information on the
neutrino mixing matrix $U$ which is now fairly
well-known \cite{snoon}. This
review is aimed at providing a short summary of the theoretical
expectations on the magnitude of one of the elements of $U$,
namely $U_{e3}$.

The $U_{e3}$ controls the strength of CP violation in neutrino
oscillations. It also leads to important sub-leading effects in   
the solar and the atmospheric neutrino oscillations. It
would therefore play an important role in deciding 
feasibility of the planned experiments \cite{snoon} which look for these
effects. It is thus important to have a rough theoretical estimate
for $|U_{e3}|$ and the existing literature is full of this \cite{rev2,bd}. 
While
it is not possible to summarize all attempts  in this short
review, we concentrate on giving basic scenarios which lead to
definite expectations on $|U_{e3}|$ illustrating them with
specific models when appropriate. The basic message which emerges
from this study is that there exists very large class of
well-motivated theoretical scenarios within which relatively large
($>0.05$) values for $\ue3$ are possible. In some specific
scenarios, the present day phenomenology provides a lower bound 
on $U_{e3}$ which is close to the existing upper bound making us
think that discovery of a non-zero $\ue3$ is just round the corner!

Elements of $U$ entering the description of neutrino oscillations
are determined by the matrices $U_l$ and
$U_\nu$ which diagonalize the charged lepton and the neutrino mass
matrices $\ml$ and $\mnu$ respectively, $U=U_l^\dagger U_\nu$. One can
always
choose the standard CKM form for $U_\nu$.
The most general $U_l$ then depends \cite{tg} on three angles
$\theta_{ijl}$, 
two Majorana  and three Dirac phases in such a way that the 
$U$ has the MNS form \cite{rev2}. The 
$U$ is thus determined by 12 different parameters in general.
 
In spite of the above complexity, one could have meaningful
predictions for $\ue3$ using ($i$) simplifying theoretical assumptions
such as quark lepton symmetry ($ii$) educated guesses following from
the existing knowledge on neutrino oscillations and/or using ($iii$)
some flavour symmetry. Unlike other angles, $\ue3$ is small;
the combined analysis of the CHOOZ and the atmospheric data
imply $\ue3\leq 0.26$ at $3\sigma$. It then makes sense to
start with neutrino mass textures which
lead to zero $\ue3$ and then use some perturbations to
obtain estimates for $\ue3$. We will systematically follow this
approach. Before doing this, let us see what are generic
expectations for $\ue3$. These expectations are based on a
natural assumptions that the charged lepton mixing angles
$\theta_{ijl}$ are
determined by the corresponding lepton mass ratios, e.g. through
square root formula. This implies
$$ \theta_{13 l}<< \theta_{12 l}\sim \sqrt{\frac{m_e}{m_\mu}}~.$$
Since $U=U_l^\dagger U_\nu$, one gets
\begin{equation} \ue3\approx |s_{13 \nu}-s_{12 l} \sin\theta_A| ~,
\label {gr} \end{equation}
where $\theta_A$ is the atmospheric mixing angle. In the absence of
any cancellations one finds that $\ue3$ is at least
$$\ue3 \approx \ord(\frac{1}{\sqrt{2}} \sqrt{\frac{m_e}
{m_\mu}})\approx 0.05 $$ which is near the projected detectable
value in the long baseline experiments with super beams \cite{snoon}.
While this generic value gets realized in several models \cite{gen,demo1},
substantial
deviations from it are possible since both $\theta_{13 \nu}$ and
$\theta_{12 l}$ contribute to
eq.(\ref{gr}) and these contributions may get added
or subtracted. 
\section{Leptonic mass structures with zero $U_{e3}$}
A systematic study of perturbations over structures which give zero $\ue3$ 
can give us some idea on
possible ranges which $|U_{e3}|$ can take. With this view in mind, we list
below various theoretical structures for the leptonic mass matrices
leading to zero $\ue3$. The zero $\ue3$ may be
attributed \cite{lop1,alt,sm} 
to specific
texture of the charged lepton mass matrix $\ml$ (with almost diagonal
$\mnu$)
or to the specific choice of $\mnu$ (with almost diagonal $\ml$). We
consider both these cases. In the latter case,
our choices for the  structures of $\mnu$ are such that the
atmospheric neutrino mixing and mass scale arise at the zeroth order
in all of them. 
\subsection{ Normal Hierarchy}
\begin{equation} ({\rm A})~~~ \mnu=m\left(
\begin{array}{ccc}0&0&0\\0&s^2&s c\\0&s
c&c^2\\ 
\end{array}
\right) \end{equation}
Here s(c) denotes the sine (cosine) of the atmospheric neutrino mixing
angle $\theta_A$ and $m^2$
corresponds to $\atm$. The solar scale and mixing angle are
absent at the zeroth order. The above structure describes the normal
neutrino mass hierarchy with $m_{\nu_3}\gg m_{\nu_{1,2}}$. The standard
seesaw picture can generate the above texture when contribution of only
one right handed neutrino dominates the seesaw mechanism \cite{srh}.

\subsection{Inverted Hierarchy}
One can write two dominant structures corresponding to the inverted mass
hierarchies ($m_{\nu_1}\approx m_{\nu_{2}}\gg m_{\nu_3}$) for the
neutrinos:
\begin{eqnarray}(B1)~ \mnu=m\left( \begin{array}{ccc}\pm 1&0&0\\0&s^2&s
c\\0&s
c&c^2\\ 
\end{array}
\right)&~~~& 
(B2)~\mnu=m\left( \begin{array}{ccc}0&c&s\\c&0&0
\\s&0&0\\ 
\end{array}
\right)  
\end{eqnarray}
Theoretically, texture $B2$ arise naturally from the $L_e-L_\mu-L_\tau$
symmetry \cite{rev2}. This structure
generates maximal solar mixing at the zeroth order although $\solm=0$
in case of both ($B1$) and ($B2$).
\subsection{$\mu-\tau$ symmetric structure}
The following is a more complex structure giving zero $\ue3$:
\begin{equation} \mnu=\left( \begin{array}{ccc}a&b&b\\b&d&e \\b&e
&d\\ 
\end{array}
\right) \label{mtau}\end{equation}
The above structure arise \cite{grim} from imposition of the $\mu-\tau$
interchange
symmetry on $\mnu$. Unlike the previous textures, the neutrino spectrum
in this case can be normal, inverted or quasi degenerate depending upon
the values of the parameters in eq.(\ref{mtau}). The atmospheric mixing
angle
is implied to be maximal while the solar mixing angle is arbitrary.
\subsection{Zero $\ue3$ from $M_l$}
There are two interesting textures for the charged leptons which
lead to zero $\ue3$ if the neutrino mass matrix is diagonal. These are
\begin{equation} \makebox{(C1) Democratic:}~~~~~ \ml=m_l\left(
\begin{array}{ccc}1&1&1\\1&1&1 \\1&1
&1\\ 
\end{array}
\right) \end{equation}
\begin{equation}\makebox{(C2) Lopsided}~~~~~ {\mathcal M}_l=m_l\left(
\begin{array}{ccc}0&0&\rho'\\0&0&\rho\\0&\e
&1\\ 
\end{array}
\right) \label{lops} \end{equation}
with $\rho\sim\rho'\sim \ord(1);\e\ll 1$. The democratic structure
 \cite{demo1,demo2} may be argued to arise from an
$S_3\times S_3$
symmetry. The lopsided \cite{lop1} structure can be nicely embedded in
grand
unified theory and has been extensively studied in this context.
There exists more examples of  structures with zero $\ue3$ 
\cite{sf}. More generally, vanishing of $\ue3$ can be understood
\cite{agt}
as a consequence of the invariance of the neutrino mass matrix in
flavour basis under a class of discrete $Z_2$ symmetries which can
in fact be used to characterize all possible textures with
zero $\ue3$. Alternative
possibility is the absence of any specific texture or symmetry in
$\mnu$. This possibility termed as anarchy \cite{anarchy}.
generally tends to give large mixing angles with a lower bound
$\geq 0.1$ on $\ue3$. 
\section{Perturbations and non-zero $\ue3$}
The above discussed basic structures can be perturbed by adding 
small parameters in $\mnu$,$\ml$ or in both.
The predicted $\ue3$ differ considerably in these two 
cases. In many cases, perturbations to $\mnu$ generate both the solar
scale and $\ue3$. This
leads to correlations
$$ \ue3\approx \sqrt{\soa}\approx 0.2 ~~ {\rm OR} ~~ \ue3\approx
\soa\approx 0.04 $$
\subsection{Perturbation to A}
The perturbed texture (A) may be written as
\begin{equation}(A')~~~~ \mnu=m\left(
\begin{array}{ccc}\e_{11}&\e_{12}&\e_{13}\\\e_{12}&s^2+\e_{22}&s
c+\e_{23}\\\e_{13}&s c+\e_{23}&c^2+\e_{33}\\ 
\end{array}
\right)~. \end{equation}
Without fine tuning or cancellations, the above structure can be 
shown \cite{bd} to lead to
$$\ue3\approx \ord(1) \sqrt{\soa}\approx 0.2 $$
Thus various models based on this simple texture tend to
predict rather large values for $\ue3$. Let us take an example
\begin{equation} \mnu=m\left(
\begin{array}{ccc}0&0&\lambda\\0&s^2&s
c\\ \lambda&s c&c^2\\ 
\end{array}
\right)~~~  \label{examp}\end{equation}
The small perturbation $\lambda$ simultaneously leads to
large solar mixing angle, solar scale and $\ue3$
 $$\tan2 \sola\sim -\frac{2 s}{\lambda c^2}~~;~~\soa\sim
2 \lambda ^3 s c^2~~;~~ \ue3\sim c\lambda ~. $$
Correlations among these imply a $\lambda$ independent relation
\begin{equation}
|U_{e3}|=\frac{\tan 2 \sola}{2 \tan\theta_A} \left(\soa \cos
2\theta_{\odot}\right)^{1/2}\approx 0.13  
\label{relation}
\end{equation}
This relation \cite{ab} is more general than its derivation presented
here.
All neutrino mass textures \cite{zero1} with $(\mnu)_{11}=(\mnu)_{12}=0$
imply\footnote{If $(\mnu)_{13}=0$ instead of $(\mnu)_{12}$ then one gets
an equivalent relation with $\tan\theta_A\rightarrow \cot\theta_A$ in
eq.(\ref{relation}).} the above relation and hence lead to quite large
$\ue3$.
Correlations between zeros in textures of the leptonic mass matrices
and $\ue3$ have been systematically studied \cite{zero1,zero3}.
Many "texture zero" analysis
presented in the literature may be regarded as perturbations to textures
(A) or (B) and lead to \cite{zero3} values of
$\ue3$ in the range $0.02-0.2$. Quite a few of them actually predict
rather large values for $\ue3$ as implied by eq.(\ref{relation}). 

The example of eq.(\ref{examp}) is realistic and arises in the minimal
$SO(10)$ model \cite{goh}
employing type II seesaw mechanism for neutrino mass generation. Because
of the quark
lepton unification, the $\lambda$ gets related to the Cabibbo angle
giving rather large $\ue3$ $\sim \frac{0.2}{\sqrt{2}}\approx 0.14$. 

There are many other examples in which $\ue3$ gets related to
the parameters in the quark sector either due to imposition of some GUT
symmetry \cite{gut1,gut2,gut3,gut4,gut5,gut6} or due to the assumed quark
lepton symmetry \cite{ql1,ql2,ql3}. The resulting predictions are given in
the Table 1.
\begin{table}[ph]
{\footnotesize
\begin{tabular}{@{}ccc@{}}
\hline
{} &{} &{} \\[-1.5ex]
{Assumptions} & $\ue3$ &Examples\\[1ex]
\hline
{} &{} &{}\\[-1.5ex]
Dominant Contribution from $U_l$ &$\ord(\sqrt{\frac{m_e}{m_\mu}})\approx
0.05$
&\cite{gen,demo1}\\[1ex]
\hline
Type-II seesaw mechanism&$\ord(\sqrt{\soa})\sim 0.2$
&\cite{goh}\\[1ex]
\hline
GUT, Family Symmetry&0.15&\cite{gut1}\\[1ex]
{}&0.014&\cite{gut2}\\[1ex]
{}&$\ord(\theta_c)$&\cite{gut3}\\[1ex]
{}&0.24&\cite{gut4}\\[1ex]
Models with two right handed neutrinos&0.07 (0.01) &\cite{gut5}(\cite{gut6})
 \\[1ex]
\hline
Quark-Lepton symmetry&0.05 &\cite{ql1}\\[1ex]
{}&0.04-0.18&\cite{ql2}\\[1ex]
{}&0.06-0.2&\cite{ql3}\\[1ex] 
\hline
Corrected Bi-Maximal $U_\nu$&$\geq 0.1$&\cite{tani}\\[1ex]
Corrected Bi-Maximal $U_l$&0.02&\cite{sm}\\[1ex]
\hline
Radiative $\solm$ in MSSM&$\simeq 0.1$&\cite{am}\\[1ex]
Radiative $\ue3$,Degenerate spectrum&$\ord(0.1)$&\cite{asj,miura}\\[1ex]
\hline
Anarchy& $\geq \ord(0.1)$&\cite{anarchy}\\[1ex]
Randomly perturbed textures&{}&{}\\[1ex]
Normal Hierarchy&$\ord(\sqrt{\soa})\sim 0.2$&{}\\[1ex]
Inverted Hierarchy&$\leq 0.01$&\cite{random}\\[1ex]
\hline
Effects beyond GUT scale& $\leq 0.04$&\cite{venya}\\[1ex]
\hline
\end{tabular}\label{table1} }
\vspace*{13pt}
\caption{Predicted $\ue3$ under different assumptions and some
illustrative
examples.}
\end{table}

\subsection{Perturbation to B2}
One needs to introduce \cite{sf} rather large perturbations to the
structure (B2)
in order to induce the solar neutrino oscillations through the LMA
solution. This requires introducing deviation of $\sola$ from the maximal
value and generation of a non-zero solar scale. The perturbations which do
this also tend to generate \cite{tani,am} rather large
$\ue3$ in many cases although there exists an example in which $B2$ 
can be perturbed to obtain the solar scale without generating
any corrections to $\ue3$ \cite{gl}.

The non-maximal $\sola$ can be generated by small leptonic mixing
angles
$\theta_{ij l}$. This also generates $\ue3$ and if  $\theta_{12 l}$
dominates as in the quark sector then one finds \cite{tani,am}
$$\tan^2 \sola=1-4 U_{e3} ~.$$
The current bound $\tan^2\sola\leq 0.64$ implies $\ue3\geq 0.09$.
Thus a large value of $\ue3$ is forced in this scenario.

The breaking of
the  $L_e-L_\mu-L_\tau$ symmetry in the charged lepton sector cannot
induce the solar scale at the tree level but it can do so
radiatively. It is found that the LMA solution cannot be
radiatively generated in supersymmetric models with zero $13$ mixing
at a high scale. One needs sizable value \cite{am} for this mixing to do
this. As
a result,
these models predict \cite{am} rather large values $\geq 0.1$ for $\ue3$.

Alternative possibility is to assume zero $\ue3$ at a high scale.
It can be generated along with the LMA solution in case of the standard
model \cite{asj} or in the presence of some seed value for $\solm$ 
\cite{miura}. The former gives small values for $\ue3$ while in the latter
case one can get $\ue3\sim 0.1$ if neutrino spectrum is quasi degenerate
and/or $\tan\beta$ is high.

\subsection{Perturbations to $C$}
Lopsided models \cite{lop1} with texture (C2) lead to
bi-large leptonic mixing. The perturbation to this
structure due to $U_\nu$  lead to a non-zero $\ue3$. If
$U_\nu$ has the CKM form then it provides \cite{sm,raidal} a nice
explanation for the empirical relation 
$\theta_\odot=\frac{\pi}{4}-\theta_C$. 
In this case one finds \cite{sm} rather small $\ue3\approx \sin\sola
|V_{cb}|\approx
0.02$.

The democratic structure for $\ml$ needs to be perturbed to obtain 
non-zero masses for the first two generations. This perturbation also
generates corrections to $U_l$ and hence a non-zero $\ue3$  which
is related to the charged lepton masses. Depending on the type of
perturbations, one can get relatively large \cite{ql1},
$\ord(\sqrt{\frac{m_e}{m_\mu}})\sim 0.07$ , or small \cite{demo2}
$\ord(\frac{m_e}{m_\mu})\sim 0.005$ $\ue3$ in this scenario.  

\section{Conclusions}
Analysis of various textures for the leptonic mass matrices giving zero
$\ue3$ shows that perturbations to these textures tend to generate
relatively large $\ue3$ in many examples although there exist
several examples with small values for $\ue3$. Summary is given in Table
1.\\ \\ 
\noindent {\bf Acknowledgment:} I am thankful to the organizers of NOON04
for giving
me an opportunity to give this talk. This work was supported by a grant
from the Japan Society for the Promotion of Science. I am thankful to
H. Minakata for hospitality and several discussions related to this work.


\begin{thebibliography}{99}

\bibitem{snoon} A. Yu. Smirnov, Summary talk at
NOON2004, Feb. 11-15, 2004.

\bibitem{rev2} For recent reviews, G. Altarelli and F. Feruglio,
{\it Phys. Rep.} {\bf 320}, 295 (1999); S. F. King {\it Rep. Prog. Phys.}
{\bf 67}, 107 (2004). 

\bibitem{bd}  S. M. Barr and I. Dorsner, {\it Nucl. Phys.} {\bf B 585},
 79 (2000). 

\bibitem{tg} C. Guinti and M. Tanimoto, {\it Phys. Rev.} {\bf D66},
113006 (2002).

\bibitem{gen} R. Barbieri, P. Creminelli and A. Romanino, {\it
Nucl. Phys.} {\bf B559}, 17 (1999); K. S. Babu, J. C. Pati and
F. Wilczek, {\it Nucl. Phys.} {\bf B566} 33 (2000); A. Ibarra and G. Ross,
{\it
Phys. Lett.} {\bf B575}, 279 (2003); G. Ross and L. Valesco-Sevilla {\it
Nucl. Phys.} {\bf B653}, 3 (2003); J. Ferrandis and S. Pakvasa,
hep-ph/0409204.

\bibitem{demo1} H. Fritzsch and Z. Xing, {\it Phys. Lett.} {\bf
B372}, 265 (1996);{\it ibid} {\bf B598} 237 (2004); M. Fukugita,
M. Tanimoto and T. Yanagida, {\it
Phys. Rev.} {\bf D57}, 4429 (1998). 

\bibitem{lop1} K. S. Babu and S. M. Barr, {\it Phys. Lett.} {\it B381}, 
202
(1992); {\it ibid} {\bf B525}, 289 (2002); J. Sato and T. Yanagida, {\it
Phys. Lett
} {\bf B430}, 127 (1998); G. Altarelli and F. Feruglio, {\it Phys. Lett
} {\bf B451}, 388 (1999).
  
\bibitem{alt} G. Altarelli, F. Feruglio and  I. Masina, hep-ph/0402155;
S. Antusch and S. F. King, hep-ph/0403053; A. Romanino, hep-ph/0402258.

\bibitem{sm} H. Minakata and A. Yu. Smirnov, hep-ph/0405088.

\bibitem{srh} A. Yu. Smirnov, {\it Phys. Rev.} {\bf D48}, 3264
(1993); S. F. King, {\it Phys. Lett.} {\bf B439}, 350 (1998).

\bibitem{grim} W. Grimus and L. Lavoura, {\it JHEP} {\bf 07}, 045
(2001); E. Ma, {\it Phys. Rev. } {\bf D66}, 117301 (2002); W. Grimus and
L. Lavoura, {\it
Phys. Lett. } {\bf B572}, 189 (2003); E. Ma and G. Rajasekaran, {\it
Phys. Rev. } {\bf D68}, 071302 (2003); R. N. Mohapatra, hep-ph/0408187.

\bibitem{demo2} E. K. Akhmedov {\it et al}, {\it Phys. Lett.} {\bf B498},
237 (2001).

\bibitem{sf} M. Frigerio and A. Yu. Smirnov, {\it Phys. Rev.} {\bf D67},
013007 (2003). 

\bibitem{agt} W. Grimus, A. S. Joshipura, S. Kaneko, 
L. Lavoura, H. Sawanaka and  M. Tanimoto, hep-ph/0408123.

\bibitem{anarchy} L. J. Hall, H. Murayama and N. Weiner, {\it
Phys. Rev. Lett.} {\bf B501}, 115 (2001); A. de Gouvea and H. Murayama,
{\it Phys. Lett.} {\bf B571}, 209 (2003).

\bibitem{ab} E.K. Akhmedov, G.C. Branco, and M.N. Rebelo,
{\it Phys. Rev. Lett.} {\bf 84}, 3535 (2000).

\bibitem{zero1} P.H. Frampton, S.L. Glashow, and D. Marfatia,
{\it Phys. Lett.} {\bf B 536}, 79 (2002); Z.-Z. Xing, {\it Phys. Lett.}
 {\bf B 530}, 159 (2002); 
W.-L. Guo and Z.-Z. Xing, {\it Phys. Rev.} {\bf D 67}, 053002 (2003);
B.R. Desai, D.P. Roy, and A.R. Vaucher, {\it Mod. Phys. Lett.} {\bf A 18},
1355 (2003); Q. Shafi and Z. Tavartkitadze, hep-ph/0401235; M. Tanimoto,
Invited talk at NOON2003 (2003), hep-ph/0305274;
Z.-Z. Xing, Invited talk at NOON2004 (2004), hep-ph/0406049.
 
 \bibitem{zero3} R. Barbieri, T. Hambye, and A. Romanino,
{\it JHEP} {\bf 0303}, 017 (2003). 

\bibitem{goh} S. H. Goh, R. N. Mohapatra and S. Ng, {\it Phys. Lett.} {\bf
B570}, 215 (2003);{\it ibid} {\it Phys. Rev.} {\bf D68} 115008
(2003); B. Bajc, G. Senjanovic and F. Vissani, {\it
Phys. Rev. Lett.} {\bf 90}, 051802 (2003).

\bibitem{gut1} M.-C. Chen and K.T. Mahanthappa, {\it Phys. Rev.} 
{\bf D68}, 17201 (2003).

\bibitem{gut2} S. M. Barr and A. Albright, {\it Phys. Rev.} {\bf D64},
073010 (2001).

\bibitem{gut3} R. Kitano and  Y. Mimura, {\it Phys. Rev.}{\bf D63},
016008 (2001); M. Bando and N. Maekawa {\it Prog. Theo. Phys.} {\bf 106},
1255 (2001).

\bibitem{gut4} S. F. King and G. Ross, hep-ph/0307190.

\bibitem{gut5} M. Raidal and A. Strumia, {\it Phys. Lett.} { \bf B553}, 72
(2003).

\bibitem{gut6} S. Raby, {\it Phys. Lett.} {\bf B561}, 119 (2003).

\bibitem{ql1}  Y. Koide, {\it Phys. Rev. } {\bf D69}, 093001 (2004).

\bibitem{ql2} M. Fukugita, M. Tanimoto and T. Yanagida, {\it
Phys. Lett.} {\bf B562}, 273 (2003).  

\bibitem{ql3} I. Dorsner and A. Yu. Smirnov, hep-ph/0403305.

\bibitem{tani} C. Giunti and M. Tanimoto, {\it Phys. Rev.} {\bf D66}, 
053013 (2002); P. H. Frampton, S. T. Petcov and
W. Rodejohann, hep-ph/0401206; W. Rodejohann, hep-ph/0403236.
  
\bibitem{am} A. S. Joshipura and S. Mohanty, {\it Phys. Rev.} {\bf D67},
091302 (2003).

\bibitem{gl} W. Grimus and L. Lavoura, hep-ph/0410279.

\bibitem{asj} Anjan S. Joshipura, {\it Phys. Lett.} {\bf 543}, 276 (2002);
Anjan S. Joshipura and S. D. Rindani, {\it Phys. Rev.} {\bf D67}, 073009
(2003); Anjan S. Joshipura, N. Singh and S. D. Rindani, {\it Nucl. Phys.}
{\bf B660}, 362 (2003).

\bibitem{miura} T. Miura, T. Shindou nd E. Takasugi, {\it Phys. Rev.} {\bf
D66}, 093002 (2002).

\bibitem{raidal} M. Raidal, hep-ph/0404046.

\bibitem{random} A. De. Gouvea, hep-ph/0401220; F. Vissani, {\it 
Phys. Lett.} {\bf B508} 79 (2001).  

\bibitem{venya} F. Vissani, M. Narayan and V. Berezinsky, {\it
Phys. Lett.} {\bf B571}, 209 (2003).
 
\end{thebibliography}
\end{document}